\documentclass{aa}
\usepackage[dvips]{graphicx}
\usepackage{psfig}

\newcommand{\lapp}{\mbox{\raisebox{-0.3em}{$\stackrel{\textstyle <}{\sim}$}}}
\newcommand{\gapp}{\mbox{\raisebox{-0.3em}{$\stackrel{\textstyle >}{\sim}$}}}

\begin{document}

\title{B1524$-$136: A CSS quasar with two-sided radio jets}

\author{F. Mantovani \inst{1} \and D.J. Saikia \inst{2} \and M. Bondi \inst{1}
\and W. Junor \inst{3} \and C.J. Salter \inst{4} \and R. Ricci \inst{5} 
}

\titlerunning{A CSS quasar with two-sided radio jets}

\offprints{F. Mantovani}

\institute{
Istituto di Radioastronomia, CNR, Via P. Gobetti 101, Bologna, Italy 
\and
National Centre for Radio Astrophysics, TIFR, Post Bag 3, Ganeshkhind, Pune 411 007, India
\and
Department of Physics and Astronomy, University of New Mexico, 800 Yale Blvd., N.E., 
Albuquerque, NM 87131, USA 
\and
Arecibo Observatory, HC3 Box 53995, Arecibo, Puerto Rico PR 00612, USA
\and
Scuola Internazionale Superiore di Studi Avanzati, Via Beirut 4, 34014 Trieste, Italy
}

\date{Received 0000; Accepted 0000}

\maketitle

\begin{abstract}
We present MERLIN, global VLBI and VLBA observations of the high-luminosity, compact 
steep-spectrum 
quasar B1524$-$136 at cm wavelengths. These observations reveal well-defined radio jets on 
both sides of the active nucleus, a situation which is almost unique amongst high-luminosity 
radio quasars. However, the radio jets on opposite sides are very dissimilar, and the overall
radio structure appears highly distorted. We discuss possible implications of these
observations.
\end{abstract}


\section{Introduction }
Although most high-luminosity, extragalactic radio sources belonging to Fanaroff-Riley
class II (FRII) have two reasonably symmetric lobes of
radio emission on opposite sides of the parent optical object, the radio jets
in these sources are overwhelmingly one-sided (cf. Bridle \& Perley  1984).
This is particularly true for quasars where radio jets are detected much more
frequently than in radio galaxies. For example,
in a study of 12 extended 3CR quasars with the Very Large Array (VLA), Bridle
et al. (1994) made deep images with typical rms noises of 20$\mu$Jy/beam. They
detected jets in all 12 sources and candidate counter-jets in 7.
However, there is no unambiguous counter-jet in any of these sources.
In a similar study of a sample of 13 radio galaxies by Fernini et al. (1993, 1997), 
only two have radio jets and neither of these show evidence of any counter jet.
Although radio jets are not frequently detected in FRII radio galaxies,
two of the good examples of counter-jets seen in FRII radio sources
are associated with relatively nearby radio galaxies.  
One of these is the well-studied classical double-lobed source, 
Cygnus A, which is at a redshift of 0.056. A jet towards
the north-west from the nucleus and a faint counter-jet  towards the south-east
can be seen on both parsec and kiloparsec scales along a position angle, 
PA$=$284$\pm$2$^\circ$
(Perley, Dreher \& Cowan 1984; Linfield 1985; Carilli et al. 1996; Krichbaum,
Alef \& Witzel 1996; Sorathia et al. 1996).  Another FRII radio galaxy where a 
well-collimated jet and counter-jet have been detected is 3C353, which is also 
at a relatively small redshift of 0.03 (Swain, Bridle \& Baum 1998). 

Radio jets are believed to be signatures of the beams carrying energy
from the nucleus to the outer lobes, and there have been a number of suggestions
to explain the apparent asymmetry of the jets despite the outer lobes being
reasonably symmetric. These include the possibilities that 
(i) the jets are intrinsically symmetric
but appear one-sided or highly asymmetric because the brightness of the receding
one is diminished due to bulk relativistic motion while the approaching jet is
boosted for the same reason (Rees 1978; Blandford \&
K\"{o}nigl 1979; Scheuer \& Readhead 1979); (ii) the active beam on the
counter-jet side has a low synchrotron emissivity and (iii) the energy supply is 
intrinsically asymmetric, but supplies energy alternately on opposite sides of
the nucleus (cf. Saikia \& Wiita 1982; Rudnick \& Edgar 1984).

There appears to be a consensus of opinion that the observed asymmetry or
one-sidedness of the jets is due to bulk relativistic motion. Although
early arguments relied on the strong correlation of jet detection with 
core strength (cf. Saikia 1984) and the parsec- and kiloparsec-scale jets
being on the same side (cf. Bridle \& Perley 1984), the strongest evidence comes 
from the correlation
of depolarization asymmetry with jet sidedness (Laing 1988; Garrington et al.
1988; Garrington, Conway \& Leahy 1991). This would also  be consistent
with jets being seen more frequently in quasars and also being very asymmetric
and one-sided in these, since quasars are believed to be inclined at smaller
angles to the line of sight than radio galaxies (cf. Barthel 1989).

In this paper, using observations made with milliarcsec resolution, we report the 
detection  of two-sided jets in a compact steep-spectrum (CSS) quasar. The CSSs are defined to
be of subgalactic dimensions, with an overall size $\lapp$20 kpc and a steep high-frequency
spectrum having $\alpha\geq$0.5, where S$\propto\nu^{-\alpha}$.  They are
believed to be young sources at an early stage of evolution (Fanti et al. 1995;
Readhead et al. 1996a,b; O'Dea 1998; Owsianik \& Conway 1998). We present below
the results of our observations of the CSS quasar B1524$-$136 and discuss their 
implications.

\subsection{B1524$-$136}
The radio source B1524$-$136 is associated with a quasar at a redshift of 
1.687, so that 1$^{\prime\prime}$ corresponds to 8.44 kpc in a Universe with H$_o$=
50 km s$^{-1}$ Mpc$^{-1}$ and q$_o$=0.5. It has a steep radio
spectrum ($\alpha^{5000}_{750}$ $\approx$ 0.64) which appears to flatten below a 
few hundred MHz
(cf. Steppe et al. 1995). It is a well-known low-frequency variable source
(Bondi et al. 1996 and references therein), although it does not exhibit
significant variations at cm wavelengths (Padrielli et al. 1987). 
VLA observations at 5 GHz find B1524$-$136 
to be slightly extended, while at 15 GHz it is resolved into a double source
(Mantovani et al. 1994). Interplanetary scintillation observations at 327 MHz
show that about 50 per cent of the total flux density is from a compact structure with a
size of $\lapp$100 milliarcsec (Jeyakumar et al. 2000). 

\section{Radio observations}

\noindent {\bf MERLIN}
The source was observed with MERLIN on 1995 May 06 at 5 GHz for $\approx$8 hr using
a bandwidth of 28 MHz. Initial
values for the telescope and correlator gains were determined from a short
observation of the bright unresolved calibrator B0552+398 (DA193). The primary
flux density and polarization calibrator was 3C286 and the secondary calibrator
was B1511$+$100. 

\begin{figure}
\vbox{
  \psfig{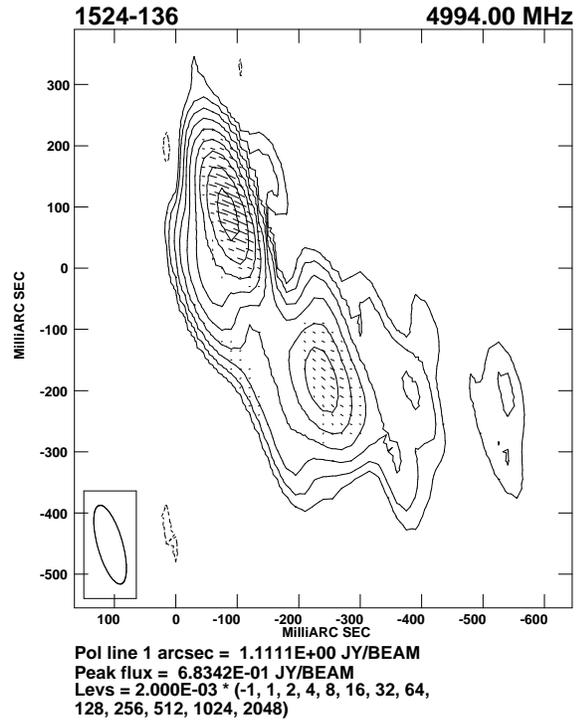}
   }
\caption[]{The The MERLIN 5-GHz image with an angular resolution 
of 133$\times$41 mas along a PA of 16$^\circ$. The total-intensity contours are
overlayed by vectors representing the linearly polarized intensity. The 
vector orientations give the polarization position angle. 
}
\end{figure}

\vspace{0.2cm}
\noindent {\bf Global VLBI MK2}
The global VLBI MK2 observations were made on 1992 September 21 with a central
frequency of 1663.99 MHz and a 2-MHz bandwidth. The array consisted of antennas
at Onsala, Effelsberg, Westerbork, Jodrell Bank (the Lovell telescope), Medicina, 
Green Bank, Haystack, 
VLBA-KP, VLBA-PT, VLBA-LA, VLBA-NL, VLBA-BR and VLBA-OV. The source was
tracked for a total of 7.5 hr in a  snap-shot observing programme. The data
were processed at the Caltech Block2 correlator.

\vspace{0.2cm}
\noindent {\bf VLBA}
These observations were carried out on 1996 Feb 04 with the Very Long Baseline Array (VLBA) and
one antenna of the VLA recording both right- and left-circular polarization
(RCP, LCP) using one-bit sampling. The source was observed at 8.4 GHz for 4.3 hr with a 
bandwidth of 32 MHz. Amplitude calibration was derived
from measurements of the system temperature during the observations and the
known antenna gain of each element of the array. Complex correlation coefficients
were recovered at the Array Operations Center in Socorro. Phase gradients in 
frequency and time were corrected by the use of global fringe fitting (Schwab
\& Cotton 1983). 

\begin{figure}
\vbox{
  \psfig{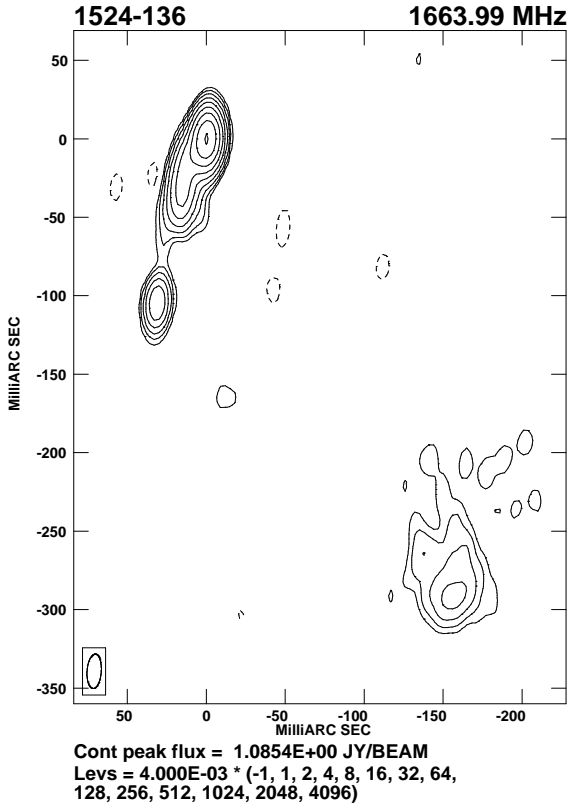}
   }
\caption[]{The global MK2 VLBI image at 1.7 GHz with an angular resolution of 22$\times$9
mas along a PA of $-$4$^\circ$.
}
\end{figure}

\section{The radio images}

Earlier VLA A-array observations of B1524$-$136 at 5 GHz show the source to be
slightly extended, while the 15-GHz observations show it to
be clearly double-lobed with an angular size of 0.32 arcsec along a PA of 30$^\circ$. 
The weaker southern component is 4.3 per cent polarized
at 15 GHz compared to 2.5 per cent for the northern component (Mantovani et al. 1994). 
Recent polarization observations with the VLBA show the northern jet and hot-spot
to have very high rotation measures (RMs) in the range of $-$3000 to $-$10,000 rad m$^{-2}$
in the rest frame of the source (Mantovani et al. 2002). 

The MERLIN 5-GHz image with an angular resolution of 133$\times$41 mas along a 
PA of 16$^\circ$ and an rms noise of 0.3 mJy/beam is shown in Fig. 1. This image
is of higher resolution than the VLA images and  
reveals greater details of the structure. The two main components, northern
and southern, are clearly seen and are 4.4 and 7.9 per cent polarized
respectively. The peak brightness of the northern component is higher 
than the southern one by a factor of about 7.
In addition, there is a barely-resolved component south of the northern component
which does not show signficant polarized emission, the degree of polarization being
$\lapp$1 per cent.  This suggests that this component is likely to be the nucleus of the
active galaxy. 

The global MK2 VLBI image at 1.7 GHz (Fig. 2), with an angular resolution of 22$\times$9
mas along a PA of $-$4$^\circ$ and an rms noise of 1.3 mJy/beam, reveals the core 
clearly, making B1524$-$136 a highly non-collinear radio source.  The supplement of 
the angle formed at the core by 
the outer hotspots is 62$^\circ$.  Such distorted structures are not uncommon among 
CSS objects (cf. Saikia et al.  1995, 2001) and are perhaps caused by deflection of 
one or both jets by dense clouds of gas. However, intrinsic distortions may appear amplified
by projection effects since the object is associated with a quasar and may thus be inclined
at a relatively small angle to the line of sight. The peak and total core flux density 
estimated by fitting a two-dimensional Gaussian are 128 mJy/beam and 140 mJy respectively. 
At this resolution, the northern hot-spot is brighter than the southern one 
by  a factor of about 25. In addition to the core and the hot-spots there is 
evidence of a jet which connects the core to the northern hotspot. By fitting two-dimensional
Gaussians to the hot-spot and jet, and taking slices across the jet, we estimate the
width of the jet to be about 7 mas and its length to be about 35 mas. This qualifies
it to be called a jet using the definition of Bridle \& Perley (1984). There is,
however, a gap or region of very low emission between the core and the beginning of
the radio jet. The hot-spot has a size of about 7 mas and an integrated flux density
of 1400 mJy, and it must be the dominant component contributing to the scintillating
flux density at 327 MHz (Jeyakumar et al. 2000) and low-frequency variability
at 408 MHz (Bondi et al. 1996).

\begin{figure}
\vbox{
  \psfig{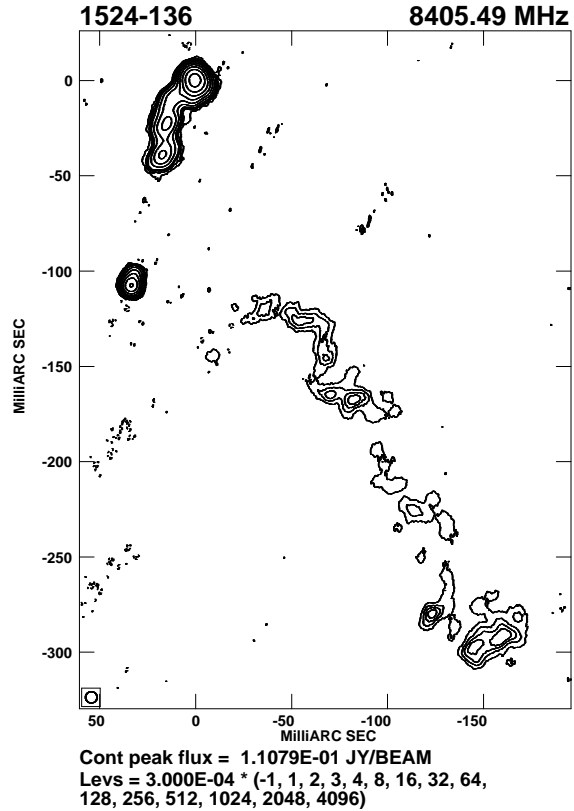}
   }
\caption[]{The VLBA image at 8.4 GHz restored with a circular beam of 6 mas.
}
\end{figure}

The VLBA image at 8.4 GHz has the highest resolution of our images, 6 mas,
and is also the most sensitive one with an
rms noise of 100 $\mu$Jy/beam. It shows a couple of prominent knots in the jet, which 
appears to bend by about 40$^\circ$ before entering the 
hot-spot (Fig. 3). The core is clearly extended to the north along the direction of the jet. 
A two-dimensional Gaussian fit shows the core to have peak and total flux densities of
40 mJy/beam and 48 mJy respectively, and the deconvolved size to be about 4 mas along
a PA=165$^\circ$. The core is unresolved in the orthogonal direction. The 
peak flux density of the core at 1.7 GHz and its total flux density at 8.4 GHz yields a 
spectral index of 0.6. However, the spectrum is likely to be flatter than this because of
the coarser resolution at the lower frequency and the presence of a jet-like structure
close to the nucleus. There is a 
gap of emission of about 50 mas between the nuclear jet and the beginning of the more 
extended one which connects to the dominant hot-spot. 

The most striking feature of the VLBA image is the dectection of a well-defined 
counter-jet which exhibits some three cycles of oscillation before entering the 
southern hot-spot. Here too
there appears to be a gap of about 60 mas between the core and the beginning of the
counter-jet.  This is 
perhaps the clearest detection yet of a jet and a counter-jet in a quasar. The mean
brightness of the knots in the counter-jet is weaker than the main jet by a factor
of about 11, while the peak brightness of the jet-side hot-spot is brighter than the 
one on the counter-jet side by a factor of about 70.  

\section{Discussion}

The most striking result of these observations of B1524$-$136 is the detection of both 
a jet and a counter-jet in a quasar. Assuming that the unified scheme for powerful 
radio galaxies and 
quasars is broadly correct (cf. Barthel 1989), the object should be inclined at less
than about 45$^\circ$ to the line of sight. The fraction of emission from the radio
core, f$_c$, at an emitted frequency of 8 GHz is 0.09, assuming a spectral index of 
0 for the core and 0.7 for the extended emission. The equivalent value of R, 
the ratio of the core to the extended flux density, is 0.10. Adopting the
values of R$_T$=0.0008 and $\gamma$=10, which give a reasonable fit to the observed 
distribution of R in FRII 3CR sources (cf. Saikia \& Kulkarni 1994), the observed value 
of R for B1524$-$136 implies an angle of inclination to the line of sight, 
$\theta\sim$20$^\circ$. 
This value does not change significantly even if we assume the core spectral index
to be 0.5. Here R$_T$ is the value of R transverse to the line-of-sight 
and $\gamma$ is
the bulk Lorentz factor. The ratio of the mean brightness of the knots in the jet to 
those in the
counter-jet is 11,  which implies a bulk velocity for the jet of $\approx$0.5c if the apparent
brightness asymmetry is entirely due to relativistic beaming. This velocity estimate 
increases by
only about 15 per cent even if the counter-jet is bent away from the jet axis by as much
as 40$^\circ$. 

The jetted side is closer to the nucleus, which is not consistent with expectations
of a symmetric external environment (cf. Saikia 1981; Scheuer 1995). This hot-spot
being closer and brighter is possibly due to a higher dissipation of energy on the
jetted side due to interaction with a much denser external medium on the northern 
side. Although we do not have information to determine the RM of the southern hot-spot,
evidence of such a dense medium on the northern side can be seen in the RM of this 
region which ranges from $-$3000 to $-$10000 rad m$^{-2}$ in the rest frame of the
source (Mantovani et al. 2002).
The brightness asymmetry of the hot-spots in the VLBA image is higher than that of
the jets by a factor of about 6. Even if the hot-spots are travelling close to 
the speed of the jets, the additional asymmetry in brightness would require 
the density in the northern side to be higher than the southern one by about 40 
using a model for the propagation of jets in an asymmetric environment
(Eilek \& Shore 1989; Gopal-Krishna \& Wiita 1991; Saikia et al. 1996). 

The oscillations seen in the counter-jet are reminescent of the development of
Kelvin-Helmholtz instabilities (cf. Hardee 1987). For a helical oscillation
(n=1), we can estimate the Mach number of the jet using the relation for the
fastest growing oscillation or resonant wavelength given by 
$\lambda_h/r \sim 5.2M/(0.66+\eta^{1/2})$ where $\lambda_h$ is the wavelength of
the helical oscillations, $r$ is the jet radius, $M$ is the Mach number and 
$\eta$ is the ratio of the jet density to the density of the external medium.
Estimating the average jet radius to be 3.5 mas by fitting two-dimensional Gaussians to 
the brighter knots in the counter-jet, we find the Mach number
to be 2.5 for a light jet ($\eta\rightarrow$0) and 6.3 in the highly entrained
limit ($\eta\rightarrow$1).

The gaps in emission between the core and both the jet and counter-jet could be
due to quasi-continuous activity in the nucleus. The gap of about 50 mas, 
a jet speed of 0.5c and an angle of inclination of 20$^\circ$ indicates a time 
scale of $\approx$10$^4$ yr for this 
quasi-continuous activity. This discontinuity is on a much smaller time scale 
than the repeated
encounters with external galaxies or mergers which may completely re-ignite radio
activity in the nucleus  leading to the formation of double-double radio sources.
The typical time scale for these events is several Myr (cf. Schoenmakers et al. 2000,
and references therein). 

Although we have considered just a simple scenario above to
explain the large-scale properties of the source, the situation could
be far more complicated. The large projected misalignment between the
two jets, plus the oscillations in the counter-jet, suggest that
the bending is three-dimensional, and that different sections of both
jets could be inclined at different angles to the line of sight. Also,
the jets could decelerate as they traverse outwards from the
nucleus, and the degree of deceleration could be different on opposite
sides of the nucleus. These would all affect the observed brightness
ratio of the jets.

Considering the main jet, its PA is aligned within a few degrees of the
extension seen in the core and it is reasonable to assume that the jet
is straight till it bends by about 40$^\circ$ before entering the hotspot.
For a jet traversing outwards with a velocity $v=\beta c$,
corresponding to a Lorentz factor, $\gamma$, the observed flux density
of even the approaching jet will be Doppler diminished when
$\gamma(1-\beta \cos\theta)>$1. As noted earlier, the 3CRR radio
sources suggest a bulk Lorentz factor, $\gamma\sim$10.  The observed flux 
density of the approaching jet would then be Doppler diminished for an angle of
inclination $\theta\gapp$25$^\circ$; the corresponding brightness ratio being 
$>$2000. A deceleration of the jets to about 0.5c by the region where the jets
are first detected would result in a decrease in Doppler dimming and might 
provide a viable explanation of the existence of gaps and the detection
of the radio jets at about 50 mas from the nucleus.

We can test this scenario further by examining whether it provides a 
satisfactory explanation to the ratios of the brightnesses of the outer 
jets to about 3 times the rms noise in the gaps of emission. The observed 
ratio is $\gapp$35 for the approaching jet, and $\gapp$3 for the 
counter-jet. Assuming the jets to be intrinsically of reasonably uniform brightness 
along their lengths, the velocities discussed earlier would require the angle of
inclination to be about 45$^\circ$, which is the dividing line between radio galaxies
and quasars in the unified scheme (Barthel 1989). However, although this is
consistent with the unified scheme, the observed value of core prominence would
imply that the intrinsic value, R$_T$ would have to be about 20 times greater than
the canonical value of 0.0008 mentioned above. It is therefore conceivable that the 
intrinsic brightness of the jet is lower in the gaps, and this could be further 
diminished by the effects of relativistic beaming.

A plausible scenario is one in which the jet is inclined at about 25$^\circ$ to
the line of sight, and the counter-jet is bent by about 50$^\circ$ such that it
is about 75$^\circ$ to the line of sight.
This will decrease the degree of Doppler dimming, making the
counter-jet visible, and will also help explain the large observed misalignment of 
the source. The velocity required to explain the brightness asymmetry of the jets
is then $\approx$0.6c. Although this does not exhaust the range of scenarios that
might explain the situation, it provides a reasonable explanation of the large-scale
features of the source. A deeper image to detect radio emission in the gaps 
between the nucleus and the jets, and proper motion studies of the knots should
enable us to build a better model for the source.  
     
\section{Concluding remarks}
We report the detection of a well-defined radio jet and a counter-jet in the compact 
steep-spectrum quasar B1524$-$136, which is almost unique in a radio quasar.
We present MERLIN, global VLBI and VLBA observations of the source.
The radio source is highly non-collinear and asymmetric, and was reported recently 
to have a high RM of several thousand rad m$^{-2}$ on the northern side. An angle
of inclination of about 25$^\circ$, a jet speed of about 0.6c, an intrinsic misalignment
of about 50$^\circ$, such that the counter-jet is at about 75$^\circ$ to the line of
sight, and a density asymmetry on opposite sides of the nucleus by a factor of 
$\approx$40 provides a consistent explanation of the observed large-scale properties 
of the source.

\begin{acknowledgements}
We thank an anonymous referee for his valuable suggestions to widen the scope of
the discussions. 
The Very Long Baseline Array and the Very Large Array are facilities of 
the National Radio Astronomy Observatory, USA, 
operated by Associated Universities, Inc., for the National Science Foundation. MERLIN is a UK
national facility operated by the University of Manchester on behalf of PPARC. We thank
the staff of the different observatories which made these observations possible.
We have made use of the NASA/IPAC Extragalactic Database (NED), which is operated by the 
Jet Propulsion Laboratory, California Institute of Technology under contract with the 
National Aeronautics and Space Administration. 
\end{acknowledgements}

\end{document}